\documentclass[fleqn]{revtex4}
\usepackage{braket}
\usepackage{blindtext}
\usepackage{amsmath}
\usepackage{graphicx}
\usepackage{multirow}

\begin{document}

\title{NRQED approach to the fine and hyperfine structure corrections of order $m\alpha^6$ and $m\alpha^6(m/M)$ - Application to the hydrogen atom}

\author{M Haidar$^1$, Z-X Zhong$^2$, V I Korobov$^3$ and J-Ph~Karr$^{1,4}$}

\affiliation{$^1$Laboratoire Kastler Brossel, Sorbonne Universit\'e, CNRS, ENS-PSL Research University, Coll\`ege de France, 4 place Jussieu, F-75005 Paris, France}
\affiliation{$^2$Division of Theoretical and Interdisciplinary Research, State Key Laboratory of Magnetic Resonance and Atomic and Molecular Physics, Wuhan Institute of Physics and Mathematics, Chinese Academy of Sciences, Wuhan 430071, China}
\affiliation{$^3$Bogoliubov Laboratory of Theoretical Physics, Joint Institute for Nuclear Research, Dubna 141980, Russia}
\affiliation{$^4$Universit\'e d'Evry-Val d'Essonne, Universit\'e Paris-Saclay, Boulevard Fran\c cois Mitterrand, F-91000 Evry, France}

\begin{abstract}
The NRQED approach is applied to the calculation of relativistic corrections to the fine and hyperfine structure of hydrogenlike atoms at orders $m\alpha^6$ and $m\alpha^6(m/M)$. Results are found to be in agreement with those of the relativistic theory. This confirms that the derived NRQED effective potentials are correct, and may be used for studying more complex atoms or molecules. Furthermore, we verify the equivalence between different forms of the NRQED Lagrangian used in the literature.
\end{abstract}

\maketitle

\section{Introduction}

Precision spectroscopy of simple atoms and molecules is a fruitful approach for testing fundamental physics at a low-energy scale. Since the discovery of the Lamb shift in hydrogen, comparisons between experiments and predictions of the bound-state QED theory have been performed at ever increased levels of accuracy, as experimental progress stimulated development of theoretical methods to compute high-order QED corrections. Among these, the nonrelativistic quantum electrodynamics (NRQED) approach~\cite{Caswell1986,Kinoshita1996} is a powerful tool to study QED corrections in weakly bound (low-$Z$) few-body systems. It has been applied to hydrogenlike (two-body) systems: muonium~\cite{Kinoshita1996,Nio1997} positronium~\cite{Pachucki1997,Czarnecki1999,Adkins2018} and the hydrogen atom~\cite{Jentschura2005}, but also to three-body systems such as the helium atom~\cite{Pachucki1998,Patkos2016} and hydrogen molecular ions~\cite{Korobov2017,Zhong2018}, and to four-body systems like Li, Be$^+$~\cite{Puchalski2014,Puchalski2015} or the hydrogen molecule~\cite{Puchalski2019}, to cite only a few examples.

Here, we use NRQED to calculate relativistic corrections at the $m\alpha^6$ and $m\alpha^6(m/M)$ orders, more specifically, those contributing to the fine and hyperfine splitting. This is motivated by recent experimental advances in the HD$^+$ molecular ion, where the comparison with theory is currently limited by the hyperfine structure calculations~\cite{Alighanbari2018,Patra2019}. So far, the hyperfine coefficients have been calculated at the leading orders $m\alpha^4$ and $m\alpha^5$ using the Breit-Pauli Hamiltonian with account of the anomalous magnetic moment~\cite{Bakalov2006,Korobov2006}. Higher-order corrections have been included only for the leading term i.e. the electron-nucleus spin-spin Fermi interaction~\cite{Korobov2009,Korobov2016}. This led us to derive the complete effective Hamiltonian for the electron spin-orbit and electron-nucleus spin-spin interactions in hydrogen molecular ions (HMI) (H$_2^+$, HD$^+$, and their isotopes)~\cite{Korobov2019}.

The NRQED approach consists in constructing from QED a nonrelativistic Lagrangian describing the interaction of an electron (or a spin-1/2 nucleus) with the electromagnetic field, and then using it to calculate the QED corrections by applying the nonrelativistic perturbation theory. The NRQED Lagrangian may be constructed {\em ab initio} by writing all possible interactions satisfying the required symmetries; its coefficient are then fixed by imposing that the NRQED and QED scattering amplitudes coincide up to the desired order~\cite{Kinoshita1996,Hill2013}. This procedure leads to a unique, gauge-invariant expression of the Lagrangian, which we have used in our work on the hyperfine structure of HMI. Alternatively, one can obtain the NRQED Hamiltonian directly from the Dirac Hamiltonian through Foldy-Wouthuysen (FW) transformations~\cite{Pachucki2005}. In this case, the expression of the effective Hamiltonian is not uniquely defined, and the form used e.g. in recent works on $m\alpha^6(m/M)$-order corrections to the spin-averaged energy levels in helium~\cite{Patkos2016} and HMI~\cite{Zhong2018} differs from the gauge-invariant form.

The hydrogen atom, where the exact fine and hyperfine splitting in the nonrecoil limit is known from the relativistic theory (see e.g.~\cite{Jentschura2006} for a summary of results on the hyperfine structure), plays an essential role to cross-check the derivation of the NRQED effective Hamiltonian. In the present work, we derive the effective Hamiltonian at the $m\alpha^6$ and $m\alpha^6(m/M)$ orders describing spin-dependent interactions in a hydrogen atom, using both forms of the NRQED Hamiltonian discussed above. We have used the same notations as in Ref.~\cite{Korobov2019} where the corresponding terms are derived in the case of HMI. The effective Hamiltonian is then used to calculate the complete fine and hyperfine structure corrections for the $2P$ state, which are found to coincide with the $(Z\alpha)$-expansion of relativistic results. This shows the equivalence between the two forms of the NRQED Hamiltonian, while the operators appearing in the effective Hamiltonian are different. This work may also serve as an introduction to the use of NRQED for calculation of higher-order relativistic corrections.

Natural relativistic units are used in Secs.~\ref{notations}-\ref{2nd-order}. For application to the $2P$ state (Sec.~\ref{H-2P}) we switch to atomic units.

\section{Notations} \label{notations}

In the NRQED framework, the general expression of the correction to the energy levels at order $m\alpha^6$ is
\begin{equation}
\Delta E^{(6)} = \left\langle \psi \middle| H^{(4)} Q (E_0 - H_0)^{-1} Q H^{(4)} \middle| \psi \right\rangle + \left\langle \psi \middle| H^{(6)} \middle| \psi \right\rangle \label{alpha6}
\end{equation}
where $H_0$, $E_0$, and $\psi$ are respectively the nonrelativistic (Schr\"odinger) Hamiltonian, energy, and wave function. One takes into account the finite nuclear mass $M$:
\begin{equation} \label{Hamiltonian}
H_0 = \frac{\mathbf{P}^2}{2M}+\frac{\mathbf{p}_e^2}{2m} + V = \frac{\mathbf{p}^2}{2m_r} + V,
\end{equation}
where $\mathbf{p} = \mathbf{p_e} = -\mathbf{P}$, $V=-\frac{Z\alpha}{r}$, and $m_r=mM/(m+M)$. $Q$ is a projection operator on a subspace orthogonal to $\psi$, and $H^{(4)}$ is the Breit-Pauli Hamiltonian yielding the leading-order ($m\alpha^4$) relativistic correction. Since our goal is to calculate the $m\alpha^6$ and $m\alpha^6(m/M)$ orders, we select the terms of orders $m\alpha^4$ and $m\alpha^4(m/M)$:
\begin{equation}
H^{(4)} = H_B + H_{rec} + H_{so} + H_{so-M} + H_{ss}^{(0)} + H_{ss}^{(2)} + H_{so-N}, \label{Breit-Pauli}
\end{equation}
\begin{equation}
\begin{array}{@{}l}\displaystyle
H_B = -\frac{p^4}{8m^3} + \frac{\pi Z\alpha}{2 m^2} \, \delta(\mathbf{r}), \label{p4-darwin}
\\[3mm]\displaystyle
H_{rec} = \frac{Z\alpha}{2}\,\frac{p^i}{m}\left(\frac{\delta^{ij}}{r}+\frac{r^ir^j}{r^3}\right)\frac{P^j}{M},
\\[3mm]\displaystyle
H_{so} =
\frac{Z\alpha}{2m^2}\,\frac{[\mathbf{r}\times\mathbf{p}]}{r^3}\,\mathbf{s}_e,
\\[3mm]\displaystyle
H_{so-M} =
-\frac{Z\alpha}{mM}\,\frac{[\mathbf{r}\times\mathbf{P}]}{r^3}\,\mathbf{s}_e,
\\[3mm]\displaystyle
H_{ss}^{(0)} =
- \frac{8\pi}{3}\boldsymbol{\mu}_e\boldsymbol{\mu}_M\,\delta(\mathbf{r}),
\\[3mm]\displaystyle
H_{ss}^{(2)} =
\frac{\boldsymbol{\mu}_e\boldsymbol{\mu}_M}{r^3}
-3\frac{(\boldsymbol{\mu}_e\mathbf{r})(\boldsymbol{\mu}_M\mathbf{r})}{r^5},
\\[3mm]\displaystyle
H_{so-N} =
\frac{\alpha}{m}\,\frac{[\mathbf{r}\times\mathbf{p}]}{r^3}\,\frac{\boldsymbol{\mu}_M}{|e|}.
\end{array}
\end{equation}
Here, $\boldsymbol{\mu}_e$ and $\boldsymbol{\mu}_M$ are respectively the electronic and nuclear magnetic moments, which may be expressed in terms of the electronic and nuclear spins:
\[
\boldsymbol{\mu}_e = -\frac{|e|}{m} \mathbf{s}_e  \hspace{1cm}   \boldsymbol{\mu}_M = \mu_M \, \frac{|e|}{2m_p} \frac{\mathbf{I}}{I}.
\]
For a $^1$H atom, $I = 1/2$ and $\mu_M = \mu_p = 2.79...$. Throughout the paper, $e$ denotes the electron's charge (and is thus negative), the elementary charge is then $|e|$. Note that the electron's anomalous magnetic moment is not taken into account here. The derivation of the $m\alpha^6$-order effective Hamiltonian $H^{(6)}$ appearing in the second term of Eq.~(\ref{alpha6}) is the object of Secs.~\ref{lagrangian} and~\ref{heff6}.

It should be noted that $\Delta E^{(6)}$ as written in Eq.~(\ref{alpha6}) contains contributions at all orders $m\alpha^6(m/M)^n$, $n=0,1,2\ldots$ not only because of the recoil terms present in $H^{(4)}$ and $H^{(6)}$, but also because $H_0$, $E_0$ and $\psi$, which depend on the reduced mass $m_r$, may be expanded in powers of $(m/M)$.

\section{NRQED Lagrangian} \label{lagrangian}

As discussed in the Introduction, we have used two different expressions of the NRQED Lagrangian in order to derive the effective Hamiltonian at orders $m\alpha^6$ and $m\alpha^6(m/M)$. The general form of the NRQED Lagrangian for an electron is
\begin{equation}
\mathcal{L} = \psi^*\left(i\partial_t-H\right)\psi + \mathcal{L}_{\rm contact}
\end{equation}
where $\psi$ is the two-component Pauli spinor field for an electron, and $\mathcal{L}_{\rm contact}$ represents the contact type interactions. Since the latter do not contribute to the quantities of interest here (note that contact terms vanish for a state of angular momentum $l \neq 0$), they will not be considered further.

\subsection{Foldy-Wouthuysen-Pachucki Hamiltonian}

One way of deriving the NRQED Hamiltonian is to use successive FW transformations of the Dirac Hamiltonian as done in several papers by Pachucki and co-workers~\cite{Patkos2016,Pachucki2005,Pachucki2006}. We will use as our starting point Eq.~(23) of Ref.~\cite{Patkos2016}:
\begin{equation}\label{FW}
\begin{array}{@{}l}\displaystyle
H_{\rm FWP} = eA_0 + \frac{\boldsymbol{\pi}^2}{2m} -\frac{e}{2m} \, \boldsymbol{\sigma}\!\cdot\!\mathbf{B}
\\[3mm]\hspace{12.5mm}\displaystyle
-\frac{e}{8m^2} \, (\boldsymbol{\nabla}\!\cdot\!\mathbf{E}_{\parallel})
+\frac{e^2}{2m^2}\,\boldsymbol{\sigma}\!\cdot\!(\mathbf{E}_{\parallel}\!\times\!\mathbf{A})
-\frac{e}{8m^2}\,\boldsymbol{\sigma}\!\cdot\!(\mathbf{E}_{\parallel}\!\times\!\mathbf{p} - \mathbf{p}\!\times\!\mathbf{E}_{\parallel})
\\[3mm]\hspace{12.5mm}\displaystyle
 - \frac{\boldsymbol{\pi}^4}{8m^3} +\frac{e^2}{8m^3}\mathbf{E}^2_{\parallel}
+\frac{e}{8m^3}\Bigl\{p^2,\boldsymbol{\sigma}\!\cdot\!\mathbf{B}\Bigr\}
-\frac{ie}{16m^3}
\Bigl[\boldsymbol{\sigma}\!\cdot\!
\bigl(\mathbf{p}\!\times\!\mathbf{A}\!-\!\mathbf{A}\!\times\!\mathbf{p}\bigr), p^2
\Bigr]
\\[3mm]\hspace{12.5mm}\displaystyle
+\frac{5e}{128m^4}\left[ p^2, [ p^2, A_0] \right]
-\frac{3e}{64m^4}\,\left\{p^2,(\nabla^2 A_0)\right\}
+\frac{3e}{32m^4}\Bigl\{p^2,\boldsymbol{\sigma}\!\cdot\!(\mathbf{E}_{\parallel}\!\times\!\mathbf{p})\Bigr\}
+\frac{p^6}{16m^5} \,,
\end{array}
\end{equation}
where $\boldsymbol{\pi}=\mathbf{p}\!-\!e\mathbf{A}$, , $\mathbf{E}=-\partial_t\mathbf{A}\!-\!\boldsymbol{\nabla}A_0$, $\mathbf{B} = \boldsymbol{\nabla}\!\times\!\mathbf{A}$ and  $\mathbf{E}_{\parallel} = -\!\boldsymbol{\nabla}\boldsymbol {A_0}$. The $\boldsymbol{\nabla}$ and $\nabla^2$ operators only act inside the parentheses that surround them.

\subsection{Gauge invariant Hamiltonian}

Alternatively, one can build the NRQED Lagrangian following an {\em ab initio} approach as initially proposed by Caswell and Lepage~\cite{Caswell1986,Kinoshita1996,Hill2013}. Starting from Eq.~(1) of Ref.~\cite{Hill2013}, and neglecting the dependence of coefficients on the anomalous magnetic moment, we obtain a gauge invariant NRQED Hamiltonian in the following form:
\begin{equation} \label{KH}
\begin{array}{@{}l}\displaystyle
H_{\rm GI} = eA_0 - \frac{\mathbf{D}^2}{2m} -\frac{e}{2m} \, \boldsymbol{\sigma}\!\cdot\!\mathbf{B}
\\[3mm]\hspace{12.5mm}\displaystyle
-\frac{e}{8m^2}\>\bigl(\mathbf{D}\!\cdot\!\mathbf{E}-\mathbf{E}\!\cdot\!\mathbf{D}\bigr)
-\frac{ie}{8m^2}\>\boldsymbol{\sigma}\!\cdot\!(\mathbf{D}\!\times\!\mathbf{E}-\mathbf{E}\!\times\!\mathbf{D})
\\[3mm]\hspace{12.5mm}\displaystyle
- \frac{\mathbf{D}^4}{8m^3} + \frac{e^2}{8m^3}\mathbf{E}^2 - \frac{e}{8m^3}\Bigl\{\mathbf{D}^2,\boldsymbol{\sigma}\!\cdot\!\mathbf{B}\Bigr\}
\\[3mm]\hspace{12.5mm}\displaystyle
+\frac{5e}{128m^4}\left[ \mathbf{D}^2, (\mathbf{D}\!\cdot\!\mathbf{E} + \mathbf{E}\!\cdot\!\mathbf{D}) \right]
+\frac{3e}{64m^4}\,\left\{\mathbf{D}^2, (\boldsymbol{\nabla}\!\cdot\!\mathbf{E}) \right\}
-\frac{3ie}{16m^4}\Bigl\{\mathbf{D}^2,\boldsymbol{\sigma} \!\cdot\! (\mathbf{D}\!\times\!\mathbf{E}-\mathbf{E}\!\times\!\mathbf{D}) \Bigr\}
-\frac{\mathbf{D}^6}{16m^5} \,,
\end{array}
\end{equation}
where  $\mathbf{D} = \boldsymbol{\nabla} - ie\mathbf{A} = i\boldsymbol{\pi}$. By simple algebraic transformations, and keeping only the terms of order up to $m\alpha^6$, one can get an expression that is easier to compare to the FWP Hamiltonian:
\begin{equation} \label{KH}
\begin{array}{@{}l}\displaystyle
H_{\rm GI} = eA_0 + \frac{\boldsymbol{\pi}^2}{2m} -\frac{e}{2m} \, \boldsymbol{\sigma}\!\cdot\!\mathbf{B}
\\[3mm]\hspace{12.5mm}\displaystyle
-\frac{e}{8m^2} \, (\boldsymbol{\nabla}\!\cdot\!\mathbf{E}_{\parallel})
+\frac{e^2}{4m^2} \, \boldsymbol{\sigma}\!\cdot\!(\mathbf{E}_{\parallel}\!\times\!\mathbf{A})
-\frac{e}{8m^2}\>\boldsymbol{\sigma}\!\cdot\!(\mathbf{E}\!\times\!\mathbf{p}-\mathbf{p}\!\times\!\mathbf{E})
\\[3mm]\hspace{12.5mm}\displaystyle
- \frac{\boldsymbol{\pi}^4}{8m^3} + \frac{e^2}{8m^3}\mathbf{E}_{\parallel}^2 + \frac{e}{8m^3}\Bigl\{p^2,\boldsymbol{\sigma}\!\cdot\!\mathbf{B}\Bigr\}
\\[3mm]\hspace{12.5mm}\displaystyle
+\frac{5e}{128m^4}\left[ p^2, [ p^2, A_0] \right]
-\frac{3e}{64m^4}\,\left\{p^2, (\nabla^2 A_0) \right\}
+\frac{3e}{32m^4}\Bigl\{p^2,\boldsymbol{\sigma}\!\cdot\!(\mathbf{E}_{\parallel}\!\times\!\mathbf{p})\Bigr\}
+\frac{p^6}{16m^5} \,.
\end{array}
\end{equation}
This expression coincides with that obtained in the penultimate step of the FW transformations leading to Eq.~(\ref{FW}), see Eqs.~(19) and (20) of Ref.~\cite{Patkos2016}. The FWP Hamiltonian~(\ref{FW}) may be obtained from Eq.~(\ref{KH}) by means of the canonical transformation $e^{iS} (H - i\partial_t) e^{-iS}$, where
\begin{equation}
S = \frac{e}{8m^2}\> \boldsymbol{\sigma} \!\cdot\! (\boldsymbol{\pi} \!\times\! \mathbf{A} - \mathbf{A} \!\times\! \boldsymbol{\pi}).
\end{equation}
\subsection{Nuclear Hamiltonian}

Since we are only interested in the first order in $m/M$, the nucleus can be treated nonrelativistically, using the Hamiltonian
\begin{equation}
H_{M} = -ZeA_0 + \frac{1}{2M} \left(\mathbf{P}-Z|e|\mathbf{A}\right)^2 - \boldsymbol{\mu}_M \!\cdot\! \mathbf{B}.
\end{equation}

\subsection{NRQED Vertices}

For the derivation of effective potentials, it is convenient to translate NRQED Hamiltonian given by Eq.~(\ref{FW}) or (\ref{KH}) in terms of NRQED vertices and ''Feynman'' rules, as done in Fig.~3 of Ref.~\cite{Kinoshita1996}. The list of vertices which play a role in interactions up to the $m\alpha^6(m/M)$ order is given in Table~\ref{vertices-table}.

\begin{table}[h]
\begin{center}
\begin{tabular}{|l|c|c|}
\hline
Name~\cite{Kinoshita1996} & Foldy-Wouthuysen Hamiltonian & Gauge invariant Hamiltonian \\
\hline
1. Coulomb & \multicolumn{2}{|c|}{$eA_0$} \\[3mm]
2. Dipole  & \multicolumn{2}{|c|}{$-e\frac{\mathbf{p}'+\mathbf{p}}{2m} \mathbf{A}$} \\[3mm]
3. Fermi   & \multicolumn{2}{|c|}{$e\frac{i[\mathbf{q}\!\times\!\boldsymbol{\sigma}]}{2m} \mathbf{A}$} \\[3mm]
4. Darwin  & \multicolumn{2}{|c|}{$-e\frac{\mathbf{q}^2}{8m^2} A_0 $} \\[3mm]
5. Seagull & $e^2 \frac{i[\mathbf{q}_1\!\times\!\boldsymbol{\sigma}]}{2m^2} A_0(\mathbf{q}_1) \mathbf{A}(\mathbf{q}_2)$ & $e^2 \frac{i[\mathbf{q}_1\!\times\!\boldsymbol{\sigma}]}{4m^2} A_0(\mathbf{q}_1) \mathbf{A}(\mathbf{q}_2)$ \\[3mm]
6. Spin-orbit        & \multicolumn{2}{|c|}{$e \frac{i[\mathbf{p}'\!\times\!\mathbf{p}]\boldsymbol{\sigma}}{4m^2} A_0$} \\[3mm]
7. Time derivative   & absent & $-e \frac{iq_0 (\mathbf{p}'+\mathbf{p})\!\times\!\boldsymbol{\sigma}}{8m^2} \mathbf{A}$ \\[3mm]
8.                   & \multicolumn{2}{|c|}{$e \frac{(p'^2 + p^2)(\mathbf{p}'+\mathbf{p})}{8m^3} \mathbf{A}$} \\[3mm]
9.                   & \multicolumn{2}{|c|}{$-e^2 \frac{q_1^i q_2^i}{8m^3} A_0 (\mathbf{q}_1) A_0 (\mathbf{q}_2)$} \\[3mm]
10. Derivative Fermi & \multicolumn{2}{|c|}{$-e \frac{i (p'^2 + p^2)(\mathbf{q} \!\times\! \boldsymbol{\sigma})}{8m^3} \mathbf{A}$} \\[3mm]
11.                  & $-e \frac{i (p'^2 - p^2) (\mathbf{p}'+\mathbf{p})\!\times\!\boldsymbol{\sigma}}{16m^3} \mathbf{A}$ & absent \\[3mm]
12.                  & \multicolumn{2}{|c|}{$ e \left( \frac{3 \mathbf{q}^2 (p'^2+p^2)}{64m^4} + \frac{5(p'^2 - p^2)^2}{128m^4} \right) A_0 $} \\[3mm]
13.                  & \multicolumn{2}{|c|}{$ -e \left( \frac {3i(p'^2 + p^2) [\mathbf{q} \!\times\! \mathbf{p}] \boldsymbol{\sigma}}{32m^4} \right) A_0 $} \\[3mm]
\hline
1M. Coulomb          & \multicolumn{2}{|c|}{$-ZeA_0$} \\[3mm]
2M. Dipole           & \multicolumn{2}{|c|}{$Ze\frac{\mathbf{P}'+\mathbf{P}}{2M} \mathbf{A}$} \\[3mm]
3M. Fermi            & \multicolumn{2}{|c|}{$i[\mathbf{Q}\!\times\!\boldsymbol{\mu}_M] \mathbf{A}$} \\[3mm]
4M. $\mathbf{A}\!\cdot\!\mathbf{A}$   & \multicolumn{2}{|c|}{$Z^2 e^2 \frac{\delta_{ij}}{2M} \mathbf{A}(\mathbf{q}_1) \mathbf{A}(\mathbf{q}_2) $} \\[3mm]
\hline
\end{tabular}
\end{center}
\caption{NRQED ''Feynman'' rules for vertices. In order to facilitate the comparison with Ref.~\cite{Kinoshita1996}, the names of the vertices considered in that work are given in the first column. The first part of the Table concerns the electron, and the second part deals with the nucleus. $\mathbf{q} = \mathbf{p}'-\mathbf{p}$, and $\mathbf{Q} = \mathbf{P}'-\mathbf{P}$.} \label{vertices-table}
\end{table}

The differences between alternative expressions of the effective Hamiltonian are clearly apparent in this Table. In the FWP Hamiltonian, \\
- the ''seagull'' vertex is multiplied by two; \\
- the ''time derivative'' vertex does not appear; \\
- the vertex numbered 11 appears in addition to the ''derivative Fermi'' vertex.

\section{Spin-dependent interactions at order $m\alpha^6$ and $m\alpha^6(m/M)$} \label{heff6}

From the NRQED vertices of Table~\ref{vertices-table} and the photon propagator in the Coulomb gauge, effective potentials are obtained by systematic application of the nonrelativistic Rayleigh-Schr\"odinger perturbation theory (see e.g.~\cite{Kinoshita1996,Pachucki2005,Pachucki2006,Korobov2009}). We will not write the explicit derivation of all terms but give some illustrative examples (see~\cite{Korobov2019} for more details).

\subsection{Coulomb photon exchange} \label{U1b}

The only spin-dependent contribution of order $m\alpha^6$ from Coulomb photon exchange is obtained by having the nucleus interact via the Coulomb vertex (1M) while the electron interacts via the higher-order vertex (13). The corresponding potential in momentum space is given by
\begin{equation}
U_{1b} = \left[ -e \frac {3i(p'^2 + p^2) [\mathbf{q} \!\times\! \mathbf{p}] \!\cdot\! \boldsymbol{\sigma}}{32m^4} \right] \left[ -Ze \right] \left[ \frac{1}{\mathbf{q}^2} \right]
\end{equation}
In order to facilitate comparison with Ref.~\cite{Korobov2019}, which deals with the HMI case, we have used the same labeling of effective potentials. A Fourier transform yields the effective potential in coordinate space,
\begin{equation}
\mathcal{U}_{1b} = - \frac{3Z\alpha}{16m^4} \left\{ p^2, \frac{1}{r^3} [\mathbf{r} \!\times\! \mathbf{p}] \!\cdot\! \mathbf{s_e} \right\}.
\end{equation}

\subsection{Transverse photon exchange without retardation}

For illustration, let us write the potential obtained by having the nucleus interact via the dipole vertex (2M), while the electron interacts via the Fermi derivative vertex (10). The potential in momentum space is
\begin{eqnarray}
U_{2b} &=& \left[ -e \frac{i (p'^2 + p^2)(\mathbf{q} \!\times\! \boldsymbol{\sigma})}{8m^3} \right] \left[ Ze\frac{\mathbf{P}'+\mathbf{P}}{2M} \right] \left[ -\frac{1}{\mathbf{q}^2} \left( \delta^{ij} - \frac{q^i q^j}{\mathbf{q}^2} \right) \right] \nonumber \\
&=& \frac{iZ\alpha}{8m^3M} (p'^2 + p^2) \frac{[\mathbf{q} \!\times\! \boldsymbol{\sigma}] \!\cdot\! \mathbf{P}}{\mathbf{q}^2}
\end{eqnarray}
After Fourier transform, one obtains
\begin{equation}
\mathcal{U}_{2b} = \frac{Z\alpha}{4m^3M} \left\{ p^2, \frac{1}{r^3} [\mathbf{r} \!\times\! \mathbf{P}] \!\cdot\! \mathbf{s_e} \right\}.
\end{equation}

\subsection{Retardation in the transverse photon exchange}

The last example we will consider in some detail is a retardation term in the exchange of one transverse photon, where the electron interacts via the time derivative vertex while the nucleus interacts via the lowest-order vertices (dipole or Fermi). The total one-photon exchange potential, which contains contributions at orders $m\alpha^5$ and above, is~\cite{Pachucki2006}
\begin{eqnarray}\label{retard}
\mathcal{U}_{3c}^{(5+)} &=&
     \int\! \frac{d^4 q}{(2\pi)^4i} \frac{4\pi}{\left( q^0 \right)^2 \!-\! \mathbf{q}^2 \!+\! i\epsilon} \left( \delta^{ij} \!-\! \frac{q^i q^j}{\mathbf{q}^2} \right) \! \left[ -\frac{ie}{8m^2} q^0 (\mathbf{p'}+\mathbf{p}) \!\times\! \boldsymbol{\sigma} \right]^i \times \nonumber \\
     && \left\{ e^{i \mathbf{q}\cdot\mathbf{r}_e} \frac{1}{E_0 - H_0 - q^0 + i\epsilon} e^{-i \mathbf{q}\cdot\mathbf{R}} \right\} \! \left( Ze \frac{\mathbf{P}' + \mathbf{P}}{2 M} \!-\! i [(-\mathbf{q})\!\times\!\boldsymbol{\mu}_M] \right)^j + (h.c.)
\end{eqnarray}
After integration over $q_0$, one gets
\begin{eqnarray}
\mathcal{U}_{3c}^{(5+)} &=& -\frac{ie}{16m^2}
\!\int\! \frac{d^3 q}{(2\pi)^3} 4\pi \left( \delta^{ij} \!-\! \frac{q^i q^j}{\mathbf{q}^2} \right) \! \left[ (\mathbf{p'} + \mathbf{p}) \!\times\! \boldsymbol{\sigma} \right]^i \times \nonumber \\
&& \left\{ e^{i \mathbf{q}\cdot\mathbf{r}_e} \frac{1}{E_0 - H_0 - q} e^{-i \mathbf{q}\cdot\mathbf{R}} \right\} \! \left( Ze \frac{\mathbf{P}}{M} \!+\! i [\mathbf{q}\!\times\!\boldsymbol{\mu}_M] \right)^j + (h.c.)
\end{eqnarray}
We perform the expansion
\begin{equation}
\frac{1}{E_0 - H_0 - q} = -\frac{1}{q} + \frac{H_0 - E_0}{q^2} - \frac{(H_0 - E_0)^2}{q^3} + \ldots
\end{equation}
where the first and second terms correspond to a contributions of order $m\alpha^5$ and $m\alpha^6$, respectively. Then,
\begin{eqnarray}
\mathcal{U}_{3c}^{(6)} &=& -\frac{ie}{16m^2}
\!\int\! \frac{d^3 q}{(2\pi)^3} \frac{4\pi}{q^2} \left( \delta^{ij} \!-\! \frac{q^i q^j}{\mathbf{q}^2} \right) \! \left[ (\mathbf{p'} + \mathbf{p}) \!\times\! \boldsymbol{\sigma} \right]^i \times \nonumber \\
&& \left\{ e^{i \mathbf{q}\cdot\mathbf{r}_e} (H_0 - E_0) e^{-i \mathbf{q}\cdot\mathbf{R}} \right\} \! \left( Ze \frac{\mathbf{P}}{M} \!+\! i [\mathbf{q}\!\times\!\boldsymbol{\mu}_M] \right)^j + (h.c.)
\end{eqnarray}
Using $\mathbf{R} = -m \mathbf{r}/(m+M)$, it is easy to show that
\begin{equation}
[ H_0 , e^{-i \mathbf{q}\cdot\mathbf{R}} ] = e^{-i \mathbf{q}\cdot\mathbf{R}} \mathcal{O}(m/M).
\end{equation}
As a consequence, neglecting a term of order $(m/M)^2$ we get
\begin{equation} \label{timederiva6}
\mathcal{U}_{3c}^{(6)} \simeq -\frac{ie}{16m^2}
\!\int\! \frac{d^3 q}{(2\pi)^3} \frac{4\pi}{q^2} \left( \delta^{ij} \!-\! \frac{q^i q^j}{\mathbf{q}^2} \right) \! \left[ (\mathbf{p'} + \mathbf{p}) \!\times\! \boldsymbol{\sigma} \right]^i e^{i \mathbf{q}\cdot\mathbf{r}} (H_0 - E_0) \! \left( Ze \frac{\mathbf{P}}{M} \!+\! i [\mathbf{q}\!\times\!\boldsymbol{\mu}_M] \right)^j + (h.c.)
\end{equation}
and since $(H_0 - E_0)$ commutes with $[\mathbf{q}\!\times\!\boldsymbol{\mu}_M]$, the nuclear spin dependent part of Eq.~(\ref{timederiva6}) has a vanishing expectation value in the state $\psi$. With the replacement $\mathbf{P} = -\mathbf{p}$ one obtains
\begin{eqnarray}
\mathcal{U}_{3c}^{(6)} &=& \frac{iZ\alpha}{16m^2M}
     \!\int\! \frac{d^3 q}{(2\pi)^3} \frac{4\pi}{q^2} \left( \delta^{ij} \!-\! \frac{q^i q^j}{\mathbf{q}^2} \right) \! \left[ (\mathbf{p}' + \mathbf{p}) \!\times\! \boldsymbol{\sigma} \right]^i e^{i \mathbf{q}\cdot\mathbf{r}} (H_0 - E_0) p^j + (h.c.) \nonumber \\
     &=& \frac{iZ\alpha}{16m^2M} \!\int\! \frac{d^3 q}{(2\pi)^3} \frac{4\pi}{q^2} \left( \delta^{ij} \!-\! \frac{q^i q^j}{\mathbf{q}^2} \right) \! \left[ (\mathbf{p}' + \mathbf{p}) \!\times\! \boldsymbol{\sigma} \right]^i e^{i \mathbf{q}\cdot\mathbf{r}} [H_0, p^j] + (h.c.)
\end{eqnarray}
After Fourier transform:
\begin{eqnarray}
\mathcal{U}_{3c}^{(6)} &=& \frac{iZ\alpha}{8m^2M} \left[ \mathbf{p} \!\times\! \boldsymbol{\sigma} \right]^i \frac{1}{2r} \left( \delta^{ij} + \frac{r^i r^j}{r^2} \right) [V, p^j] + (h.c.) \nonumber \\
&=& -\frac{Z\alpha^2}{8m^2M} \left[ \mathbf{p} \!\times\! \boldsymbol{\sigma} \right]^i \frac{1}{2r} \left( \delta^{ij} + \frac{r^i r^j}{r^2} \right) \frac{r^j}{r^3} + (h.c.) \nonumber \\
&=& -\frac{Z\alpha^2}{8m^2M} \left[ \mathbf{p} \!\times\! \boldsymbol{\sigma} \right]^i \frac{r^i}{r^4} + (h.c.) \nonumber \\
&=& -\frac{Z\alpha^2}{2m^2M} \frac{1}{r^4} \left[ \mathbf{r} \!\times\! \mathbf{p} \right] \!\cdot\! \mathbf{s}_e = -\frac{Z\alpha^2}{2m^2M} \frac{1}{r^4} \mathbf{l} \!\cdot\! \mathbf{s}_e.
\end{eqnarray}
\subsection{Total effective Hamiltonian}

We give in this Section the complete set of spin-dependent effective operators. At the (nonrecoil) $m\alpha^6$ order, there is only one term, which is the Coulomb photon exchange considered in Sec.~\ref{U1b}:
\begin{equation}
\mathcal{U}_{1b} = - \frac{3Z\alpha}{16m^4} \left\{ p^2 , \frac{1}{r^3} \mathbf{l} \!\cdot\! \mathbf{s_e} \right\} \label{nonrecoil}
\end{equation}
The $m\alpha^6(m/M)$-order (recoil) terms are listed in Table~\ref{results-h}, where we have separated the terms depending only on the electronic spin and those on the nuclear spin, which respectively contribute to the fine and hyperfine structure.					
\begin{table}[h!]
\begin{center}
\begin{tabular}{|c|c|c|c|}
\hline
Type of interaction & Vertices & Foldy-Wouthuysen Hamiltonian & Gauge invariant Hamiltonian \\
\hline
\multirow{3}{*}{Transverse photon (no retard.)} & 10-2M   & \multicolumn{2}{|c|}{\hspace{16mm}$\mathcal{U}_{2b} = -\frac{Z\alpha}{4m^3M} \left\{ p^2, \frac{1}{r^3} \mathbf{l} \!\cdot\! \mathbf{s}_e \right\} $}   \\[2mm]
                                                & 11-2M   & $\mathcal{U}'_{2b} = \frac{1}{2} \mathcal{U}_{2b} - \left( \frac{iZ\alpha}{8m^3M} p^2 \frac{1}{r^3} [\mathbf{r}\!\times\! (\mathbf{r}\!\cdot\!\mathbf{p}) \mathbf{p}]
                                                             \!\cdot\! \mathbf{s}_e  + (h.c) \right)$ & absent \\[4mm]
\multirow{3}{*}{Transverse photon (retard.)}    & 3-2M    & \multicolumn{2}{|c|}{\hspace{20mm}$\mathcal{U}_{3b} = \frac{Z^2\alpha^2}{2m^2M} \frac{1}{r^4} \mathbf{l} \!\cdot\! \mathbf{s}_e$} \\[1mm]
                                                & 7-2M    & absent & $\mathcal{U}_{3c} = -\frac{Z^2\alpha^2}{2m^2M} \frac{1}{r^4} \mathbf{l} \!\cdot\! \mathbf{s}_e$ \\[4mm]
Seagull                                         & 5-1M-2M & $\mathcal{U}_{5a}^{(FWP)} = -\frac{Z^2}{2m^2M} \frac{1}{r^4} \mathbf{l} \!\cdot\! \mathbf{s}_e$
                                                          & $\mathcal{U}_{5a}^{(GI)} = -\frac{Z^2}{4m^2M} \frac{1}{r^4} \mathbf{l} \!\cdot\! \mathbf{s}_e$ \\[1mm]
Double Coulomb photon                           & 9-1M-1M & \multicolumn{2}{|c|}{\hspace{20mm}$\mathcal{U}_{6b} = -\frac{Z^2\alpha^2}{2m^2M} \frac{1}{r^4} \mathbf{l} \!\cdot\!  \mathbf{s}_e$} \\[1mm]
\hline
\multirow{6}{*}{Transverse photon (no retard.)} & 8-3M    & \multicolumn{2}{|c|}{\hspace{16mm}$\mathcal{U}_{2c}=- \frac{\alpha\mu_M}{4m^3m_p} \left\{ p^2 , \frac{1}{r^3} \mathbf{l} \!\cdot\! \mathbf{I} \right\}$} \\[3mm]
                                                & 10-3M   & \multicolumn{2}{|c|}{$\mathcal{U}_{2d} =-\frac{\alpha\mu_M}{4m^3m_p} \left\{ p^2, \left[ \frac{8\pi}{3} \delta(\mathbf{r}) \mathbf{s}_e \!\cdot\! \mathbf{I} - \frac{r^2
                                                            \mathbf{s}_e \!\cdot\! \mathbf{I} - 3 (\mathbf{r} \mathbf{s}_e) (\mathbf{r} \mathbf{I})}{r^5} \right] \right\} $} \\[4mm]
                                                & 11-3M   & $\mathcal{U}'_{2d} = \frac{1}{2} \mathcal{U}_{2d} - \left( \frac{i\alpha\mu_M}{4m^3m_p} p^2 \frac{ (\mathbf{r} \mathbf{p}) (\mathbf{s}_e \mathbf{I}) - (\mathbf{r}
                                                             \mathbf{s}_e) (\mathbf{p} \mathbf{I})}{r^3} + (h.c.) \right)$ & absent \\[5mm]
Seagull                                         & 5-1M-3M & $\mathcal{U}_{5b}^{(FWP)} = \frac{Z\alpha\mu_M}{m^2m_p} \frac{r^2 \mathbf{s}_e \!\cdot\! \mathbf{I} - (\mathbf{r} \mathbf{s}_e) (\mathbf{r} \mathbf{I})}{r^6}$
                                                          & $\mathcal{U}_{5b}^{(GI)} = \frac{Z\alpha\mu_M}{2m^2m_p} \frac{r^2 \mathbf{s}_e \!\cdot\! \mathbf{I} - (\mathbf{r} \mathbf{s}_e) (\mathbf{r} \mathbf{I})}{r^6}$ \\[1mm]
\hline
		\end{tabular}
	\end{center}
	\caption{ Spin-dependent effective operators at order $m\alpha^6(m/M)$ for a hydrogenlike atom. The upper and lower parts respectively correspond to interactions depending on the electronic spin only, and to those depending on the nuclear spin. \label{results-h}}
\end{table}

\section{Second-order and finite-mass corrections} \label{2nd-order}
The total second-order contribution is given by the first term of Eq.~(\ref{alpha6}). Using expression~(\ref{Breit-Pauli}) of $H^{(4)}$, we pick up the terms contributing to the electronic spin-orbit interaction (fine structure) and those depending on nuclear spin (hyperfine structure). For the fine structure, we also separate the nonrecoil ($m\alpha^6$) and recoil ($m\alpha^6(m/M)$) terms.

\subsection{Electronic spin-orbit interaction}
 \begin{description}
	\item[$\bullet$]   Nonrecoil contributions
	\begin{eqnarray}
	\Delta E_{B-so}^{(2)} &=& 2 \; \langle H_B Q (E_0 - H_0)^{-1} Q H_{so} \rangle \label{B-so} \\
	\Delta E_{so-so}^{(2)} &=& \langle H_{so} Q (E_0 - H_0)^{-1} Q H_{so} \rangle \label{so-so}
	\end{eqnarray}
	Note that the Darwin term in $H_B$ (Eq.~(\ref{p4-darwin})) vanishes because we are considering $l \neq 0$ states.
	\item[$\bullet$] Recoil contributions
	\begin{eqnarray}
	\Delta E_{B-so-M}^{(2)} &=& 2 \; \langle H_B Q (E_0 - H_0)^{-1} Q H_{so-M} \rangle \label{B-SO-M} \\
	\Delta E_{rec-so}^{(2)} &=& 2 \; \langle H_{rec} Q (E_0 - H_0)^{-1} Q H_{so} \rangle \\
	\Delta E_{so-so-M}^{(2)} &=& 2 \; \langle H_{so} Q (E_0 - H_0)^{-1} Q H_{so-M} \rangle \label{so-so-M}
	\end{eqnarray}
    We also have to take into account the corrections to the nonrecoil terms, Eqs.~(\ref{nonrecoil}) induced by the finite nuclear mass in $H_0$, $E_0$, and $\psi$, to first order in $m/M$:
    \begin{equation}
    \delta_M ( \Delta E_{fs}^{(6)} ) = \delta_M(\langle \mathcal{U}_{1b} \rangle) + \delta_M(\Delta E_{B-so}^{(2)}) + \delta_M(\Delta E_{so-so}^{(2)}). \label{finite-mass}
    \end{equation}
\end{description}

\subsection{Nuclear spin dependent contributions}

The second-order terms that involve nuclear spin at the $m\alpha^6(m/M)$ order are:
\begin{eqnarray}
\Delta E_{B-ss}^{(2)} &=& 2 \; \langle H_B Q (E_0 - H_0)^{-1} Q H_{ss}^{(2)} \rangle \label{B-ss} \\
\Delta E_{B-so-N}^{(2)} &=& 2 \; \langle H_B Q (E_0 - H_0)^{-1} Q H_{so-N} \rangle \\
\Delta E_{so-ss}^{(2)} &=& 2 \; \langle H_{so} Q (E_0 - H_0)^{-1} Q H_{ss}^{(2)} \rangle \\
\Delta E_{so-so-N}^{(2)} &=& 2 \; \langle H_{so} Q (E_0 - H_0)^{-1} Q H_{so-N} \rangle \label{so-so-N}
\end{eqnarray}
Note that the scalar part of the spin-spin interaction $H_{ss}^{(0)}$ does not appear because we are considering $l \neq 0$ states.

\section{Fine and hyperfine structure of the $2P$ state} \label{H-2P}

In this Section, we calculate analytically all the first-order, second-order and finite-mass contributions for the $2P$ state of the hydrogen atom and compare with known results from the relativistic theory. No ultraviolet divergences (at $r \to 0$) appear in any of the above expressions, because of the $r$ factor in the $2P$ wavefunction. Such divergences are found in the case of $S$ states, e.g. in the $m\alpha^6$-order correction to the spin-averaged energy levels~\cite{Adkins2005}. From here on, we switch from the relativistic units to atomic units.

\subsection{Zero-order and first-order wavefunctions}

In the limit of an infinite nuclear mass, the radial wavefunction  and non-relativistic energy of the $2P$ state are expressed as
\begin{eqnarray}
\psi_{0}(r) &=& \frac{Z^{3/2}}{2\sqrt{6}} \; (Zr) \; e^{-\frac{1}{2}Zr} \label{psi0-2P} \\
E_0 &=& -\frac{Z^2}{8}. \label{E0-2P}
\end{eqnarray}
One may notice that all the second-order perturbation terms (Eqs.~(\ref{B-so}-\ref{so-so-M}) and (\ref{B-ss}-\ref{so-so-N}) depend either on $H_B$ or $H_{so}$ (see Eq.~(\ref{p4-darwin})). In order to calculate them, we introduce the first-order perturbation wave functions  ${\psi}_B^{(1)}$ and ${\psi}_{so}^{(1)}$, defined by
\begin{eqnarray}
\left( E_0-H_0 \right)\psi_B^{(1)} &=& (H_B-\left\langle H_B \right\rangle) \psi_0 \\
\left( E_0-H_0 \right)\psi_{so}^{(1)} &=& (H_{so}-\left\langle H_{so} \right\rangle) \psi_0
\end{eqnarray}
These perturbation wavefunctions may be obtained analytically. For the $2P$ state we have
\begin{eqnarray}
\psi_{B}^{(1)}(r) &=&
Z^2\left[
\frac{1}{2}-\frac{Zr}{3}\ln{Zr}
-\frac{\gamma_E\,Zr}{3}+\frac{97\,Zr}{144}-\frac{(Zr)^2}{48}
\right]\frac{Z^{3/2}}{2\sqrt{6}}\>e^{-\frac{1}{2}Zr} \label{psi1B-2P} \\
\psi_{so}^{(1)}(r) &=&
Z^2
\left[
-\frac{1}{4}+\frac{Zr}{12}\ln{Zr}+\frac{\gamma_E\,Zr}{12}
-\frac{31\,Zr}{144}+\frac{(Zr)^2}{48}
\right]\frac{Z^{3/2}}{2\sqrt{6}}\>e^{-\frac{1}{2}Zr}
 \langle \mathbf{l}\!\cdot\!\mathbf{s}_e \rangle \label{psi1SO-2P},
\end{eqnarray}
where $\gamma_E$ is the Euler-Mascheroni constant. In the case of a finite nuclear mass, the zero- and first-order wavefunctions are obtained through the replacement $Z \to (m_r/m)Z$ in Eqs.~(\ref{psi0-2P}) and (\ref{psi1B-2P}-\ref{psi1SO-2P}), and the nonrelativistic energy through multiplication of Eq.~(\ref{E0-2P}) by $(m_r/m)$.

\subsection{Nonrecoil $m\alpha^6$-order contributions to the fine structure}

The total contribution to the fine structure splitting is the sum of the first-order and second-order terms, respectively given by Eq.~(\ref{nonrecoil}) and Eqs.~(\ref{B-so}-\ref{so-so}):
\begin{equation}
\Delta E_{fs}^{(6)} = \langle \mathcal{U}_{1b} \rangle + \Delta E_{B-so}^{(2)} + \Delta E_{so-so}^{(2)}
\end{equation}
The calculations are straightforward and require no particular explanations. One obtains
\begin{eqnarray}
\langle \mathcal{U}_{1b} \rangle &=&  -2 \frac{3 Z}{16} \; \int_0^{\infty} 2\left( E_0 + \frac{Z}{r} \right) \left| \psi_0(r) \right|^2 \frac{1}{r^3} \, r^2 dr \; \langle \mathbf{l}\!\cdot\!\mathbf{s}_e \rangle = -\frac{7 Z^6}{256} \langle \; \mathbf{l}\!\cdot\!\mathbf{s}_e \rangle \\
\Delta E_{B-so}^{(2)} &=& Z \int_0^{\infty} \psi_0 (r) \, \psi_{B}^{(1)}(r) \frac{1}{r^3} r^2 dr \; \langle \mathbf{l}\!\cdot\!\mathbf{s}_e \rangle = \frac{115 Z^6}{3456} \langle \; \mathbf{l}\!\cdot\!\mathbf{s}_e \rangle \\
\Delta E_{so-so}^{(2)} &=& \frac{Z}{2} \int_0^{\infty} \psi_0 (r) \psi_{so}^{(1)}(r) \frac{1}{r^3} \, r^2 dr \; \langle (\mathbf{l}\!\cdot\!\mathbf{s}_e)^2 \rangle = -\frac{49 Z^6}{3456} \; \langle (\mathbf{l}\!\cdot\!\mathbf{s}_e)^2 \rangle = \frac{49 Z^6}{6912} \; \langle \mathbf{l}\!\cdot\!\mathbf{s}_e \rangle + \ldots
\end{eqnarray}
In the last line, we have used the fact that in the $2p_{1/2}-2p_{3/2}$ subspace, $(\mathbf{l}\!\cdot\!\mathbf{s}_e)^2 = \frac{1}{2} - \frac{1}{2} \mathbf{l}\!\cdot\!\mathbf{s}_e$, and kept only the term that contributes to the fine-structure splitting. Note that a common factor of $\alpha^4$ is omitted in all expressions. Finally,
\begin{equation}
\Delta E_{fs}^{(6)} = \frac{5 Z^6}{384} \; \langle \mathbf{l}\!\cdot\!\mathbf{s}_e \rangle
\end{equation}
which is in agreement with the $Z\alpha$-expansion of the Dirac result (see e.g. Eq.~(3.5) of Ref.~\cite{Eides}).

\subsection{Recoil $m\alpha^6(m/M)$-order contributions to the fine structure}

Let us first use the effective Hamiltonian derived from the gauge invariant NRQED Hamiltonian of Eq.~(\ref{KH}). Collecting results from the rightmost column of Table~\ref{results-h} and from Eqs.~(\ref{B-SO-M}-\ref{finite-mass}), the total $m\alpha^6(m/M)$-order contribution is
\begin{eqnarray}
\Delta E_{fs}^{(6M)} &=& \langle \mathcal{U}_{2b} \rangle \!+\! \langle \mathcal{U}_{3b} \rangle \!+\! \langle \mathcal{U}_{3c} \rangle \!+\! \langle \mathcal{U}_{5a}^{(GI)} \rangle \!+\! \langle \mathcal{U}_{6b} \rangle \!+\! \Delta E_{B\!-\!so\!-\!M}^{(2)} \!+\! \Delta E_{rec\!-\!so}^{(2)} \!+\! \Delta E_{so\!-\!so\!-\!M}^{(2)} \!+\! \delta_M ( \Delta E_{fs}^{(6)} ) \\
&=& \left( -\frac{Z}{2} \, \frac{m}{M} \left\langle p^2 \frac{1}{r^3} \right\rangle \!-\! \frac{3Z^2}{4} \, \frac{m}{M} \left\langle \frac{1}{r^4} \right\rangle \right) \langle \mathbf{l}\!\cdot\!\mathbf{s}_e \rangle \!+\! \Delta E_{B\!-\!so\!-\!M}^{(2)} \!+\! \Delta E_{rec\!-\!so}^{(2)} \!+\! \Delta E_{so\!-\!so\!-\!M}^{(2)} \!+\! \delta_M ( \Delta E_{fs}^{(6)} ) \label{GI-fs-results}
\end{eqnarray}
Like in the preceding paragraph, a common factor of $\alpha^4$ will be omitted in all the expressions. For the first-order terms we have:
\begin{eqnarray}
\left\langle p^2 \frac{1}{r^3} \right\rangle &=& \int_0^{\infty} 2\left( E_0 + \frac{Z}{r} \right) \left| \psi_0(r) \right|^2 \frac{1}{r^3} r^2 dr = \frac{7 Z^5}{96} \\
\left\langle \frac{1}{r^4} \right\rangle &=& \int_0^{\infty} \left| \psi_0(r) \right|^2 \frac{1}{r^4} r^2 dr = \frac{Z^4}{24}
\end{eqnarray}
The second-order terms are:
\begin{eqnarray}
\Delta E_{B-so-M}^{(2)} &=& 2 \frac{m}{M} \Delta E_{B-so}^{(2)} = \frac{115 Z^6}{1728} \, \frac{m}{M} \; \langle \mathbf{l}\!\cdot\!\mathbf{s}_e \rangle \\
\Delta E_{rec-so}^{(2)} &=& -Z\frac{m}{M} \left\{ \int_0^{\infty} \frac{1}{r} 2\left( E_0 + \frac{Z}{r} \right) \psi_0 (r) \psi_{so}^{(1)}(r) r^2 dr \right. \label{rec-so} \\
&&\left. + \int_0^{\infty} \frac{1}{r^3} r \frac{\partial \psi_0}{\partial r} \psi_{so}^{(1)}(r) r^2 dr - \int_0^{\infty} \frac{1}{r^3} r \frac{\partial}{\partial r} \left( r \frac{\partial \psi_0} {\partial r} \right) \psi_{so}^{(1)}(r) r^2 dr \right\} \; \langle \mathbf{l}\!\cdot\!\mathbf{s}_e \rangle \nonumber \\
&=& \frac{35 Z^6}{576} \, \frac{m}{M} \; \langle \mathbf{l}\!\cdot\!\mathbf{s}_e \rangle \\
\Delta E_{so-so-M}^{(2)} &=& 4 \frac{m}{M} \Delta E_{so-so}^{(2)} = \frac{49 Z^6}{1728} \, \frac{m}{M} \; \langle \mathbf{l}\!\cdot\!\mathbf{s}_e \rangle
\end{eqnarray}
To get the second line of Eq.~(\ref{rec-so}) we have used $\mathbf{r} (\mathbf{r}\!\cdot\!\mathbf{p}) \mathbf{p} = (i + \mathbf{r}\!\cdot\!\mathbf{p}) \mathbf{r}\!\cdot\!\mathbf{p}$. The finite mass corrections are
\begin{eqnarray}
\delta_M(\langle \mathcal{U}_{1b} \rangle) &=& -5\frac{m}{M} \langle \mathcal{U}_{1b} \rangle = \frac{35 Z^6}{256} \, \frac{m}{M} \; \langle \mathbf{l}\!\cdot\!\mathbf{s}_e \rangle \\
\delta_M(\Delta E_{B-so}^{(2)}) &=& -6\frac{m}{M} \Delta E_{B-so}^{(2)} = -\frac{115 Z^6}{576} \, \frac{m}{M} \; \langle \mathbf{l}\!\cdot\!\mathbf{s}_e \rangle \\
\delta_M(\Delta E_{so-so}^{(2)}) &=& -5\frac{m}{M} \Delta E_{so-so}^{(2)} = -\frac{245 Z^6}{6912} \, \frac{m}{M} \; \langle \mathbf{l}\!\cdot\!\mathbf{s}_e \rangle
\end{eqnarray}
and the total finite mass correction is
\begin{equation}
\delta_M ( \Delta E_{fs}^{(6)} ) = - \frac{85 Z^6}{864} \, \frac{m}{M} \; \langle \mathbf{l}\!\cdot\!\mathbf{s}_e \rangle.
\end{equation}
Finally, the total contribution of order $Z\alpha^6(m/M)$ is
\begin{equation}
\Delta E_{fs}^{(6M)} = -\frac{Z^6}{96} \, \frac{m}{M} \; \langle \mathbf{l}\!\cdot\!\mathbf{s}_e \rangle,
\end{equation}
in agreement with the expansion in powers of $Z\alpha$ and $m/M$ of the relativistic result (Eq.~(3.5) of~\cite{Eides}).

We should now check that by using the FWP effective Hamiltonian of Eq.~(\ref{FW}) we arrive at the same result. The second-order and finite-mass terms are unchanged, and the first-order contribution becomes
\begin{eqnarray}
\Delta E_{fs-1^{st} order}^{(6M)(FWP)} &=& \langle \mathcal{U}_{2b} \rangle \!+\! \langle \mathcal{U}'_{2b} \rangle \!+\! \langle \mathcal{U}_{3b} \rangle \!+\! \langle \mathcal{U}_{5a}^{(FWP)} \rangle \!+\! \langle \mathcal{U}_{6b} \rangle \nonumber \\
&=& \left( -\frac{3Z}{4} \, \frac{m}{M} \left\langle p^2 \frac{1}{r^3} \right\rangle \!-\! \frac{Z^2}{2} \, \frac{m}{M} \left\langle \frac{1}{r^4} \right\rangle \right) \langle \mathbf{l}\!\cdot\!\mathbf{s}_e \rangle - \frac{iZ}{4} \, \frac{m}{M} \left\langle p^2 \frac{1}{r^3} [\mathbf{r}\!\times\! (\mathbf{r}\!\cdot\!\mathbf{p}) \mathbf{p}] \!\cdot\! \mathbf{s}_e \right\rangle \nonumber \\
&=& \left( -\frac{Z}{2} \, \frac{m}{M} \left\langle p^2 \frac{1}{r^3} \right\rangle \!-\! \frac{iZ}{4} \, \frac{m}{M} \left\langle p^2 \frac{1}{r^3} (\mathbf{r}\!\cdot\!\mathbf{p}) \right\rangle \!-\! \frac{Z^2}{2} \, \frac{m}{M} \left\langle \frac{1}{r^4} \right\rangle \right) \langle \mathbf{l}\!\cdot\!\mathbf{s}_e \rangle \label{FW-fs-result}
\end{eqnarray}
To get the last line, we have used the relationship $\mathbf{r}\!\times\! (\mathbf{r}\!\cdot\!\mathbf{p}) \mathbf{p} = (i + \mathbf{r}\!\cdot\!\mathbf{p}) [\mathbf{r}\!\times\!\mathbf{p}]$. Comparing Eq.~(\ref{FW-fs-result}) with the first term of Eq.~(\ref{GI-fs-results}) one can see that both results are equivalent if the equality
\begin{equation}
\left\langle i p^2 \frac{1}{r^3} (\mathbf{r}\!\cdot\!\mathbf{p}) \right\rangle = \left\langle \frac{Z}{r^4} \right\rangle \label{equiv}
\end{equation}
is verified. Using the relationship $\mathbf{r}\!\cdot\!\mathbf{p} = \frac{1}{i} r \frac{\partial}{\partial r}$ and integration by parts, it is straightforward to obtain this equality. This verifies the equivalence of results obtained from the Foldy-Wouthuysen and gauge-invariant forms of the NRQED effective Hamiltonian for an arbitrary bound state of a hydrogenlike atom.

\subsection{$m\alpha^6(m/M)$-order contributions to the hyperfine structure}

\subsubsection{Results from relativistic theory}

We recall the relativistic expression of the hyperfine energy for the $(n,l,j,F)$ level of a hydrogenlike atom in natural relativistic units~\cite{Jentschura2006,Wundt2011}:
\begin{equation}
E_{hfs}(n,l,j,F) = \alpha(Z\alpha)^3 m (2\mu_M) \frac {m}{m_p} \frac{\kappa [ 2 \kappa (\gamma + n - |\kappa|) - N ]}{N^4 \left( \kappa^2 - \frac{1}{4} \right) \gamma (4 \gamma^2 - 1)} \; \langle \mathbf{I} \!\cdot\! \mathbf{j} \rangle
\end{equation}
where $\mathbf{j} = \mathbf{l} + \mathbf{s}_e$, $\mathbf{F} = \mathbf{j} + \mathbf{I}$, $\kappa = (-1)^{j-l+\frac{1}{2}} \left( j +\frac{1}{2} \right)$ is the Dirac angular quantum number, $\gamma = \sqrt{\kappa^2 - (Z\alpha)^2}$, and $N = \sqrt{(n - |\kappa|)^2 + 2(n - |\kappa|)\gamma + \kappa^2}$ is the effective principal quantum number. Expansion of this formula in powers of $Z\alpha$ yields the relativistic correction of order $m\alpha^6(m/M)$ to the hyperfine structure~\cite{Wundt2011}:
\begin{equation}
\Delta E_{hfs}^{rel} = (Z\alpha)^2 \left[ \frac{12 \kappa^2 - 1}{2 \kappa^2 (2\kappa - 1) (2\kappa + 1)} + \frac{3}{2n} \frac{1}{|\kappa|} + \frac{3 - 8\kappa}{2n^2 ( 2\kappa - 1)} \right] E_F,
\end{equation}
where
\begin{equation}
E_F = \alpha (2 \mu_M) \frac{m}{m_p} \frac{\kappa}{|\kappa|} \frac{(Z\alpha)^3 m}{n^3 (2\kappa+1) \left(\kappa^2 - \frac{1}{4} \right)} \; \langle \mathbf{I} \!\cdot\! \mathbf{j} \rangle
\end{equation}
is the Fermi energy. For the $2P$ state one obtains, going back to atomic units:
\begin{eqnarray}
\Delta E_{hfs}^{rel} (2P_{1/2},F) &=& \frac{47}{24} (Z\alpha)^2 E_F (2P_{1/2},F) \label{hfs-rel-1} \\
E_F (2P_{1/2},F) &=& Z^3 \alpha^2 \mu_M \frac{m}{m_p} \frac{1}{9} \; \langle \mathbf{I} \!\cdot\! \mathbf{j} \rangle \\
\Delta E_{hfs}^{rel} (2P_{3/2},F) &=& \frac{7}{24} (Z\alpha)^2 E_F (2P_{3/2},F) \\
E_F (2P_{3/2},F) &=& Z^3 \alpha^2 \mu_M \frac{m}{m_p} \frac{1}{45} \; \langle \mathbf{I} \!\cdot\! \mathbf{j} \rangle \label{hfs-rel-4}
\end{eqnarray}
\subsubsection{NRQED calculation}

We will now evaluate this correction from NRQED using the gauge-invariant effective Hamiltonian~(\ref{KH}). Collecting the results from Table~\ref{results-h} and Eqs.~(\ref{B-ss}-\ref{so-so-N}) we have
\begin{equation}
\Delta E_{hfs}^{(6M)} = \langle \mathcal{U}_{2c} \rangle \!+\! \langle \mathcal{U}_{2d} \rangle \!+\! \langle \mathcal{U}_{5b}^{(GI)} \rangle \!+\! \Delta E_{B\!-\!ss}^{(2)} \!+\! \Delta E_{B\!-\!so\!-\!N}^{(2)} \!+\! \Delta E_{so\!-\!ss}^{(2)} \!+\! \Delta E_{so\!-\!so\!-\!N}^{(2)} \label{hfs-total}
\end{equation}
Various combinations of spin operators appear in the above expression, and in order to make a comparison with Eqs.~(\ref{hfs-rel-1}-\ref{hfs-rel-4}) they should be ''projected'' into $\mathbf{I} \!\cdot\! \mathbf{j}$. This is done in the Appendix~\ref{annex-spin} for all the relevant operators. We now evaluate all terms and, using the results of the Appendix~\ref{annex-spin}, express them in terms of $\mathbf{I} \!\cdot\! \mathbf{j}$. In order to alleviate the expressions, we have omitted a common factor of $\alpha^4 \mu_M (m/m_p)$.
\begin{description}
\item[$\bullet$] First-order terms:
\begin{eqnarray}
\langle \mathcal{U}_{2c} \rangle &=&  -\frac{1}{2} \left\langle p^2 \frac{1}{r^3} \right\rangle \langle \mathbf{l}\!\cdot\!\mathbf{I} \rangle = -\frac{7 Z^5}{192} \, \langle \mathbf{l}\!\cdot\!\mathbf{I} \rangle = -\frac{7 Z^5}{192} \, \frac{j(j+1)+2-3/4}{2j(j+1)} \langle \mathbf{I}\cdot\mathbf{j} \rangle \\
\langle \mathcal{U}_{2d} \rangle &=& \frac{1}{2} \left\langle p^2 \frac{1}{r^3} \right\rangle \left( \langle \mathbf{s}_e\!\cdot\!\mathbf{I} \rangle -3 \left\langle \frac{(\mathbf{r} \mathbf{s}_e) (\mathbf{r} \mathbf{I})}{r^2} \right\rangle \right) = \frac{7 Z^5}{192} \, \left( \langle \mathbf{s}_e\!\cdot\!\mathbf{I} \rangle -3 \left\langle \frac{(\mathbf{r} \mathbf{s}_e) (\mathbf{r} \mathbf{I})}{r^2} \right\rangle \right) \nonumber \\
&=& \frac{7 Z^5}{192} \, \frac{j(j+1)-2-3/4}{2j(j+1)} \langle \mathbf{I}\cdot\mathbf{j} \rangle \\
\langle \mathcal{U}_{5b}^{(GI)} \rangle &=& \frac{Z}{2} \left\langle \frac{1}{r^4} \right\rangle \, \left( \langle \mathbf{s}_e\!\cdot\!\mathbf{I} \rangle - \left\langle \frac{(\mathbf{r} \mathbf{s}_e) (\mathbf{r} \mathbf{I})}{r^2} \right\rangle \right) = \frac{Z^5}{48} \, \left( \langle \mathbf{s}_e\!\cdot\!\mathbf{I} \rangle - \left\langle \frac{(\mathbf{r} \mathbf{s}_e) (\mathbf{r} \mathbf{I})}{r^2} \right\rangle \right) \nonumber \\
&=& \frac{Z^5}{48} \, \frac{j(j+1)+1/4-2}{2j(j+1)} \langle \mathbf{I}\cdot\mathbf{j}) \rangle
\end{eqnarray}
\item[$\bullet$] Second-order terms (we recall that the first-order wavefunctions $\psi_{B}^{(1)}$ and $\psi_{so}^{(1)}$ are taken from Eqs.~(\ref{psi1B-2P}-\ref{psi1SO-2P})):
\begin{eqnarray}
\Delta E_{B-ss}^{(2)} &=& \!-\!2 \int_0^{\infty} \frac{\psi_0 (r)}{r} \psi_{B}^{(1)}(r) dr \, \left( \langle \mathbf{s}_e\!\cdot\!\mathbf{I} \rangle -3 \left\langle \frac{(\mathbf{r} \mathbf{s}_e) (\mathbf{r} \mathbf{I})}{r^2} \right\rangle \right) = -\frac{115 Z^5}{1728} \left( \langle \mathbf{s}_e\!\cdot\!\mathbf{I} \rangle -3 \left\langle \frac{(\mathbf{r} \mathbf{s}_e) (\mathbf{r} \mathbf{I})}{r^2} \right\rangle \right) \nonumber \\
&=& -\frac{115 Z^5}{1728} \, \, \frac{j(j+1)-2-3/4}{2j(j+1)} \langle \mathbf{I}\cdot\mathbf{j} \rangle \\
\Delta E_{B-so-N}^{(2)} &=& 2 \int_0^{\infty} \psi_0 (r) \psi_{B}^{(1)}(r) \frac{1}{r^3} r^2 dr \, \langle \mathbf{l}\!\cdot\!\mathbf{I}\rangle = \frac{115 Z^5}{1728} \langle \mathbf{l}\!\cdot\!\mathbf{I}\rangle = \frac{115 Z^5}{1728} \frac{j(j+1)+2-3/4}{2j(j+1)} \, \langle \mathbf{I}\!\cdot\!\mathbf{j} \rangle \\
\Delta E_{so-ss}^{(2)} &=&  -Z \int_0^{\infty} \psi_0 (r) \psi_{so}^{(1)}(r) \frac{1}{r^3} r^2 dr \, \left( \langle (\mathbf{l}\cdot\mathbf{s}_e)(\mathbf{s}_e\cdot\mathbf{I}) \rangle - 3 \left\langle \frac{(\mathbf{r}\cdot\mathbf{s}_e)(\mathbf{r}\cdot\mathbf{I})}{r^2} (\mathbf{l}\!\cdot\!\mathbf{s}_e) \right\rangle \right) \nonumber \\
&=& \frac{49 Z^5}{864} \, \left( \langle (\mathbf{l}\cdot\mathbf{s}_e)(\mathbf{s}_e\cdot\mathbf{I}) \rangle - 3 \left\langle \frac{(\mathbf{r}\cdot\mathbf{s}_e)(\mathbf{r}\cdot\mathbf{I})}{r^2} (\mathbf{l}\!\cdot\!\mathbf{s}_e) \right\rangle \right) \nonumber \\
&=& -\frac{49 Z^5}{864} \frac{j(j+1) - 4 - 3/4}{4j(j+1)} \, \langle \mathbf{I}\cdot\mathbf{j} \rangle \\
\Delta E_{so-so-N}^{(2)} &=&  Z \int_0^{\infty} \psi_0 (r) \psi_{so}^{(1)}(r) \frac{1}{r^3} r^2 dr \, \langle (\mathbf{l}\cdot\mathbf{s}_e)(\mathbf{l}\cdot\mathbf{I}) \rangle = -\frac{49 Z^5}{864} \langle (\mathbf{l}\cdot\mathbf{s}_e)(\mathbf{l}\cdot\mathbf{I}) \rangle \nonumber \\
&=& -\frac{49 Z^5}{864} \left[ 2 \frac{j(j+1) - 2 - 1/4}{2} - \frac{j(j+1) -2 - 3/4}{4} \right] \frac{\langle \mathbf{I}\cdot\mathbf{j} \rangle}{j(j+1)}
\end{eqnarray}
\end{description}
Adding up these results, we find
\begin{eqnarray}
\Delta E_{hfs}^{(6M)} (2P_{1/2},F) = \frac{47Z^5}{216} \, \langle \mathbf{I}\!\cdot\!\mathbf{j} \rangle \\
\Delta E_{hfs}^{(6M)} (2P_{3/2},F) = \frac{7Z^5}{1080} \, \langle \mathbf{I}\!\cdot\!\mathbf{j} \rangle
\end{eqnarray}
in agreement with Eqs.~(\ref{hfs-rel-1}-\ref{hfs-rel-4}). Finally, one can show that the FWP effective Hamiltonian leads to the same result, not only for the $2P$ state but for any bound state, se Appendix~\ref{proof} for details.

\section{Conclusion}

In this work, we have used the NRQED approach to calculate relativistic corrections to the fine and hyperfine structure of hydrogenlike atoms. Our results are in agreement with those obtained by expanding the relativistic results in powers of $Z\alpha$ and $m/M$. This constitutes a cross-check of the validity of the effective potentials we have derived, which may then be applied to more complex systems. Such a cross-check is very useful since in this type of calculations, the probability of mistakes is increased by the relatively large number of terms. It should be noted, though, that our results cannot be considered as a {\em complete} validation of the effective potentials we have derived in the case of HMI~\cite{Korobov2019}, because a few additional terms appear which have no equivalent in the hydrogen atom case, namely the ''crossed'' seagull terms involving both nuclei.

We have also verified the equivalence of two alternative forms of the NRQED Lagrangian. The choice of one or the other is largely a matter of taste, but it is worth noticing that the additional terms that appear when one uses the FWP Hamiltonian ($\mathcal{U}'_{2b}$ and $\mathcal{U}'_{2d}$, see Table~\ref{results-h}) have the most complicated expressions. This is, of course, not an issue in the hydrogen atom case, but may give practical reasons to choose the gauge-invariant form for application to more complex systems, where matrix elements of the effective operators can only be calculated numerically.

\section*{Acknowledgements}

Z.-X.Z. acknowledges supports from the National Natural Science Foundation of China (Grants No. 91636216, No. 11974382 and No. 11474316), and from the Chinese Academy of Sciences (the Strategic Priority Research Programme,
Grant No. XDB21020200, and the YIPA program). V.I.K. acknowledges support of the Russian Foundation for Basic Research under Grant No.~19-02-00058-a. J.-Ph.K. acknowledges support as a fellow of the Institut Universitaire de France.

\appendix

\section {Expression of spin-dependent operators in terms of $\mathbf{I} \!\cdot\! \mathbf{j}$} \label{annex-spin}

The coupling scheme of angular momenta is $\mathbf{j} = \mathbf{l} + \mathbf{s}_e$, $\mathbf{F} = \mathbf{j} + \mathbf{I}$. All the expressions below are valid within a given $(n,l,j)$ subspace.
\begin{equation}
\mathbf{l}\cdot\mathbf{I} = \frac{\mathbf{l}\cdot\mathbf{j}}{j(j+1)}(\mathbf{I}\cdot\mathbf{j}) = \frac{j(j+1)+l(l+1)-3/4}{2j(j+1)} (\mathbf{I}\cdot\mathbf{j}) \,.
\end{equation}
\begin{equation}
\mathbf{s}_e\cdot\mathbf{I} = \frac{\mathbf{s}_e\cdot\mathbf{j}}{j(j+1)}(\mathbf{I}\cdot\mathbf{j}) = \frac{j(j+1)+3/4-l(l+1)}{2j(j+1)} (\mathbf{I}\cdot\mathbf{j}) \,.
\end{equation}
\begin{equation}
\frac{(\mathbf{r}\cdot\mathbf{s}_e)(\mathbf{r}\cdot\mathbf{I})}{r^2} = \frac{(\mathbf{r}\cdot\mathbf{s}_e)(\mathbf{r}\cdot\mathbf{j})}{r^2} \frac{(\mathbf{I}\cdot\mathbf{j})}{j(j+1)}
= \frac{(\mathbf{r}\cdot\mathbf{s}_e)(\mathbf{r}\cdot\mathbf{s}_e)}{r^2} \frac{(\mathbf{I}\cdot\mathbf{j})}{j(j+1)}
= \frac{1}{4} \frac{(\mathbf{I}\cdot\mathbf{j})}{j(j+1)} \,.
\end{equation}
\begin{eqnarray}
(\mathbf{l}\cdot\mathbf{s}_e)(\mathbf{s}_e\cdot\mathbf{I}) &=& (\mathbf{l}\cdot\mathbf{s}_e)(\mathbf{j}\cdot\mathbf{s}_e) \frac{(\mathbf{I}\cdot\mathbf{j})}{j(j+1)}
\nonumber\\
&=& \left[\frac{1}{4}\mathbf{l}^2-\frac{1}{2}\mathbf{l}\cdot\mathbf{s}_e+(\mathbf{l}\cdot\mathbf{s}_e)\mathbf{s}_e^2\right] \frac{(\mathbf{I}\cdot\mathbf{j})}{j(j+1)} = \frac{j(j+1) + l(l+1) -3/4}{8j(j+1)} (\mathbf{I}\cdot\mathbf{j}) \,.
\end{eqnarray}
\begin{eqnarray}
(\mathbf{l}\cdot\mathbf{s}_e)(\mathbf{l}\cdot\mathbf{I}) &=& (\mathbf{l}\cdot\mathbf{s}_e)(\mathbf{l}\cdot\mathbf{j}) \frac{(\mathbf{I}\cdot\mathbf{j})}{j(j+1)}
= [(\mathbf{l}\cdot\mathbf{s}_e)\mathbf{l}^2+\frac{1}{4}\mathbf{l}^2-\frac{1}{2}\mathbf{l}\cdot\mathbf{s}_e] \frac{(\mathbf{I}\cdot\mathbf{j})}{j(j+1)} \\
&=& \left[ l(l+1) \frac{j(j+1) - l(l+1) - 1/4}{2} - \frac{j(j+1) -l(l+1) - 3/4}{4} \right] \frac{(\mathbf{I}\cdot\mathbf{j})}{j(j+1)} \,.
\end{eqnarray}
\begin{eqnarray}
\imath(\mathbf{r}\cdot\mathbf{s}_e)(\mathbf{p}\cdot\mathbf{I}) &=& \imath(\mathbf{r}\cdot\mathbf{s}_e)(\mathbf{p}\cdot\mathbf{j}) \frac{(\mathbf{I}\cdot\mathbf{j})}{j(j+1)}
= \imath(\mathbf{r}\cdot\mathbf{s}_e)(\mathbf{p}\cdot\mathbf{s}_e) \frac{(\mathbf{I}\cdot\mathbf{j})}{j(j+1)} \nonumber \\
&=& \left[\frac{1}{4}\imath(\mathbf{r}\cdot\mathbf{p})-\frac{1}{2}\mathbf{l}\cdot\mathbf{s}_e\right] \frac{(\mathbf{I}\cdot\mathbf{j})}{j(j+1)}
= \frac{1}{4} \left[\imath(\mathbf{r}\cdot\mathbf{p})- (j(j+1) - l(l+1) -3/4) \right] \frac{(\mathbf{I}\cdot\mathbf{j})}{j(j+1)} \,.
\end{eqnarray}
%

\section{Equivalence of the FWP and Gauge-invariant Hamiltonians for the hyperfine structure} \label{proof}
If one uses the FWP Hamiltonian, the first-order contribution becomes
\begin{equation}
\Delta E_{hfs-1^{st} order}^{(6M)(FWP)} = \langle \mathcal{U}_{2c} \rangle \!+\! \langle \mathcal{U}_{2d} \rangle  \!+\! \langle \mathcal{U}'_{2d} \rangle \!+\! \langle \mathcal{U}_{5b}^{(FWP)} \rangle
\end{equation}
Comparing with the first-order terms of Eq.~(\ref{hfs-total}), one can see that both expressions are equivalent if the following equality holds:
\begin{equation}
\langle \mathcal{U}'_{2d} \rangle = \langle \mathcal{U}_{5b}^{(GI)} \rangle - \langle \mathcal{U}_{5b}^{(FWP)} \rangle = -\langle \mathcal{U}_{5b}^{(GI)} \rangle
\end{equation}
We separate $\mathcal{U}'_{2d}$ into three terms:
\begin{eqnarray}
\langle \mathcal{U}_{2d}^{'(1)} \rangle &=& \frac{1}{2} \langle \mathcal{U}_{2d} \rangle = \frac{1}{4} \left\langle p^2 \frac{1}{r^3} \right\rangle \left( \langle \mathbf{s}_e\cdot\mathbf{I} \rangle - 3 \left\langle \frac{(\mathbf{r}\cdot\mathbf{s}_e)(\mathbf{r}\cdot\mathbf{I})}{r^2} \right\rangle \right) = \left\langle p^2 \frac{1}{r^3} \right\rangle \frac{j(j+1)-l(l+1)-3/4}{8j(j+1)} \langle \mathbf{I}\cdot\mathbf{j} \rangle \\
\langle \mathcal{U}_{2d}^{'(2)} \rangle &=& -\frac{1}{2} \left\langle i p^2 \frac{1}{r^3} (\mathbf{r}\cdot\mathbf{p}) \right\rangle \langle \mathbf{s}_e \cdot \mathbf{I}\rangle = -\frac{Z}{2} \left\langle \frac{1}{r^4} \right\rangle \langle \mathbf{s}_e \cdot \mathbf{I}\rangle = -\frac{Z}{2} \left\langle \frac{1}{r^4} \right\rangle \frac{j(j+1)+3/4-l(l+1)}{2j(j+1)} \langle \mathbf{I}\cdot\mathbf{j} \rangle \\
\langle \mathcal{U}_{2d}^{'(3)} \rangle &=& \frac{1}{2} \left\langle i p^2 \frac{1}{r^3} (\mathbf{r}\cdot\mathbf{s}_e) (\mathbf{p} \cdot \mathbf{I}) \right\rangle = \left\langle p^2 \frac{1}{r^3} \left[\imath(\mathbf{r}\cdot\mathbf{p})- (j(j+1) - l(l+1) -3/4) \right] \right\rangle \frac{\langle \mathbf{I}\cdot\mathbf{j} \rangle}{8j(j+1)} \nonumber \\
&=& \left[ Z \left\langle \frac{1}{r^4} \right\rangle - \left\langle p^2 \frac{1}{r^3} \right\rangle (j(j+1) - l(l+1) -3/4) \right] \frac{\langle \mathbf{I}\cdot\mathbf{j} \rangle}{8j(j+1)} \\
-\langle \mathcal{U}_{5b}^{(GI)} \rangle &=& -\frac{Z}{2} \left\langle \frac{1}{r^4} \right\rangle \left( \langle \mathbf{s}_e\cdot\mathbf{I} \rangle - \left\langle \frac{(\mathbf{r}\cdot\mathbf{s}_e)(\mathbf{r}\cdot\mathbf{I})}{r^2} \right\rangle \right) = -\frac{Z}{2} \left\langle \frac{1}{r^4} \right\rangle \frac{j(j+1)-l(l+1)+1/4}{2j(j+1)} \langle \mathbf{I}\cdot\mathbf{j} \rangle
\end{eqnarray}
In the above derivations, we have used the relationship~(\ref{equiv}). One finally gets
\begin{equation}
\langle \mathcal{U}_{2d}^{'(1)} \rangle + \langle \mathcal{U}_{2d}^{'(2)} \rangle + \langle \mathcal{U}_{2d}^{'(3)} \rangle = -\langle \mathcal{U}_{5b}^{(GI)} \rangle,
\end{equation}
which proves that the results from both forms of the NRQED Hamiltonian are identical for any $l \neq 0$ state of a hydrogenlike atom.

\end{document}